\documentstyle[11pt,epsfig,epsf,aasms4]{article}
\newcommand{\kms}{${\rm\ km\ s}^{-1}$}

\newcommand{\pccm}{{\rm\ cm}$^{-3}$}

\newcommand{\etal}{{et al.\,}}

\def\edcomment#1{\iffalse\marginpar{\raggedright\sl#1\/}\else\relax\fi}
\begin{document}
\title{ FLIERs as stagnation knots from post-AGB winds with polar
momentum deficiency}
 \author{Wolfgang Steffen }
\affil{Instituto de Astronom\'{\i}a y Meteorolog\'{\i}a, Universidad de
Guadalajara,  Av. Vallarta 2602, 44130 Guadalajara, Jal., M\'exico }
\authoremail{wsteffen@cencar.udg.mx}
\author{Jos\'e Alberto L\'opez }
\affil{Instituto de Astronom\'{\i}a, Universidad Nacional Aut\'onoma de
M\'exico,  Apartado Postal 877, 22800 Ensenada, B.C., M\'exico}
\author{Andrew Lim}
\affil{Department of Physics and Astronomy, University College London,
Gower Street, London WC1E 6BT, UK}

\begin{abstract}
We present an alternative model for the formation of fast low-ionization 
emission regions (FLIERs) in planetary nebulae that is able to account for 
many of their attendant characteristics and circumvent the problems on the
collimation/formation mechanisms found in previous studies.  In this model,
the section of the stellar wind flowing along the symmetry axis carries less
mechanical momentum than at higher  latitudes, and temporarily develops a
concave or inverted shock geometry. The shocked ambient material is thus
refracted towards the symmetry axis, instead of away from it, and accumulates
in the concave section. The reverse is true for the outflowing stellar wind,
which  in the reverse shock is refracted away from the axis.
It surrounds the stagnation region of the bow-shock and confines the
trapped ambient gas. The latter has time to cool and is then
compressed into a dense "stagnation knot" or "stagnation jet". In the
presence of a variable stellar wind these features may eventually overrun the
expanding nebular shell and appear as detached FLIERs. We present
representative two and three-dimensional hydrodynamic simulations of the
formation and early evolution of stagnation knots and jets and compare their
dynamical properties with those of FLIERs in planetary nebulae.  

\em{Subject Headings:} hydrodynamics - ISM: jets and outflows - ISM: kinematics
and dynamics - planetary nebulae: FLIERs and stagnation knots 
 
\end{abstract}

\section{Introduction and model description}
%

%
% REVIEW OF OBSERVATIONAL LITERATURE ON FLIERs
%

FLIERs in planetary nebulae were originally identified with the structures
previously known as ansae in elliptical planetary nebulae (Aller, 1941;
Balick et al. 1993). Their peculiar characteristics are now recognized
in a  much wider variety of PNe (e.g. Guerrero, V\'azquez \& L\'opez 1999;
Goncalves et al. 2000). Their nature has resisted a consistent
explanation to date. Initially, FLIERs were considered as pairs of knots
located symmetrically with respect to the PN nucleus and characterised by
outflow radial velocities of the order of 30-50 \kms. Ionization gradients
decrease outwards from the nebular core and in often `head-tail'
morphologies are observed (see Balick et al. 1998).

However, and as pointed out by L\'opez (2000), the
concept of FLIERs has been used in recent times to encompass nearly any [N~II]
bright knot in the periphery of PN-shells found to be traveling either with
nearly null or very high radial velocity. A single model can hardly account
for all the properties observed. Symmetric ejecta, shocks, recombination
fronts or simply dynamical instabilities may form in the  nebular rim
developing dense, low ionization knots and expanding with the rim. These
effects have not always been distinguished. 

Regarding the observed
position and velocities, projection effects have to be taken into
account as well. A number of symmetric FLIERs, like the outer pair in
NGC~7009, appear to be nearly aligned with the plane of the sky.
Although they show only a few \kms\ radial velocity, the
deprojected speed is estimated by Reay \& Atherton (1985) as 160 (D/2) km/s,
with D the distance in kpc. In some sources with extended ansae or 
strings of symmetric knots a linear increase of speed with distance has been
found with values rising up to $\approx 630$~\kms (Bryce et al. 1997, Corradi
et al. 1999, O'Connor et al. 2000). Obviously, FLIERs are only part of a much
richer variety of dynamic phenomena in PNe. 

Models have proliferated with varying degrees of success in explaining the
observed properties of these structures. For example, ionization fronts (IF)
in localized dense knots and collimated flows which produce fast knots ramming
through the shell of the PN have been discussed by Balick et al. (1998).
Dopita (1997) has discussed FLIERs in terms of shocks immersed in the strongly
radiative PN environment. 

FLIERs do not usually show a physical connection
between themselves and the nebular core. This fact has to be accounted
for by any collimation model. In this regard, Frank, Balick \& Livio (1996)
discussed the case of a cooling wind that avoids fast reexpansion via a
collimation mechanism similar to the  one described by Cant\'o et al. (1988).
In this model the wind slides along the aspherical outer shock towards
the axis forming a dense, narrow jet. Alternatively, Redman \& Dyson (1999)
have presented a model in which FLIERs represent recombination fronts (RF) in
mass-loaded jets. 

On a different approach,  Garc\'{\i}a-Segura \& L\'opez
(2000) have recently developed 3D magneto- hydrodynamic models which
successfully reproduce ansae-type structures by considering a shocked stellar
wind carrying a toroidal magnetic field (see R\'o$\dot{\rm z}$yczka \& Franco,
1996) with mass-loss rates $\leq$ 10$^{-7}$ M${_\odot}$ yr$^{-1}$. Above this
limit, jet structures develop in their model.  In the MHD model, and the others
mentioned above, the material cooling to form FLIERs comes from the stellar
wind or a mixture with material from the ambient (in the case of
mass-loading). 

In this paper we elaborate on an alternative hydrodynamical model for the
formation of axial, high-velocity ansae, introduced by Steffen \& L\'opez
(2000). In this model, FLIERs are formed from {\em external} gas swept-up by a
fast, low-density wind plowing into the ambient medium of the PN, the latter
formed from the slow dense wind of the central star during its AGB phase. It
is worth  noting  that the group of elliptical nebulae with typical
FLIERS show very extended halos, such as NGC 3242 (Meaburn, L\'opez \&
Noriega-Crespo,  2000) and NGC 7009 (Moreno-Corral, de la Fuente \&
Gutierrez, 1998)

%
% PRESENTATION OF STAGNATION KNOT MODEL
%

\subsection{The stagnation knot model}
\label{model.sec}

The idea of a "stagnation knot" was used for the first time to
reproduce the large-scale structure of the giant envelope of the PN KjPn8
(Steffen \& L\'opez, 1998). We have generalized the model in such a way 
that it is able to produce symmetric high-density knots of jets from an
uncollimated stellar wind or even from precessing jets. 
In our model, we postulate that there is a fast low-density outflow from the
central object of a planetary nebula with a deficiency of momentum along the
symmetry axis of the nebula (as compared to the momentum flowing along higher
latitudes, see Figure 1a). The deficiency of momentum flux along the axis may 
arise from an interacting binary system, as described by Soker \& Rappaport
(2000). Alternatively, precessing collimated outflows or jets
produce conical outflows (Peter \& Eichler, 1995, and Lim, 2001) which also
have the required flow geometry. 

The reduced momentum flux along the axis causes the bow-shock to advance
more slowly near the axis and hence the shock becomes concave instead of convex
(as seen from outside; see Figure 2). The {\em ambient medium} passing through
the oblique regions of this section is then refracted towards the axis,
instead of away from it (see the velocity vectors on the right of Figure 1;
these have been drawn in the reference frame of the stagnation region of the
bow-shock). This converging motion is contrary to what occurs in conventional
bow-shocks where the shocked external gas has diverging streamlines. 
Note that the "concave" shape refers to a usually spherical symmetry of the
stellar wind, i.e. a somewhat flat-top or "boxy" geometry may be enough to 
produce "polar caps" similar to those seen in the "Cat's Eye" nebula
(NGC~6543), even if it is not strictly "concave".  

The reverse shock going into the fast stellar wind, has a similar shape, but
causes the fast wind to {\em diverge} (see Figure 1). Consequently, the
accumulated external material in the stagnation region is surrounded
and confined by the high pressure of the surrounding hot gas from the
diverging shocked stellar wind. The confinement allows sufficient time for the
collected external gas to cool and be compressed to a dense knot or long
jet-like feature. These are the structures we refer to as stagnation knots. 
The fast and dilute stellar wind
diverges from the axis and is too hot to cool during the required timescales
of order $10^3$ to $10^4$ years. This is an important difference
between this model and previous models.

In many cases, the symmetric ansae are seen well outside and detached from the
main envelope. In our model this occurs once the fast stellar wind stops or
reduces its power, i.e. in the case of a non-steady outflow (see Figure 1b).
Variable PN nuclear winds have been clearly shown to operate in cases like the
Stingray Nebula (Bobrowsky et al., 1998), LMC-N66 (Pe\~na et al., 1997) and Lo
4 (Werner et al. 1992). These few well documented cases indicate that variable
winds in evolving PNe may be a common condition, rather than the exception.
In our model the wind is considered to cease or decrease temporarily a few
hundred or thousand years after the formation of the PN and the expansion of
the envelope slows down. The stagnation knot has by now accumulated sufficient
mass and, given its higher density and associated specific momentum, it slows
down at a lower rate than the envelope and consequently may escape from it
and propagates into the ambient medium. The evolution of such a dense knot
propagating through a thin ambient medium has been studied in some detail by
Jones, Kang \& Tregillis (1994) using  hydrodynamic simulations and has been
also discussed by Soker \& Regev (1998) in the context of FLIERs. 

In this paper we shall concentrate on the dynamical
and kinematical properties of the formation and initial evolution of the
stagnation knots. The study of the ionization structure around the
stagnation knot, taking into account the complex shock structures and the 
photo-ionization from the central star, is out of the scope of the present
work and will be addressed in a future paper. We show here that at least three
different types of structure can be produced: first, slow ansae which remain
inside or near the rim of the main nebula; second, fast ansae which move a
relatively large distance out of the rim; third, very long, jet-like strings
with an internal linear velocity increase as a function of distance from the
source.

\section{Simulations}

%
% CODE AND PARAMETERS OF THE SIMULATIONS
%
   
The 2D-hydrodynamical simulations in axisymmetry have been calculated using
the  Coral-code with a 5-level binary adaptive grid (Raga et al. 1995).
This code solves the equations of mass, momentum, and energy conservation
using a flux-vector-splitting scheme (van Leer 1982). The non-equilibrium
cooling as described in Biro, Raga, \& Cant\'o (1995) has been used. For low
temperatures, energy loss from the collisional excitation of [O~I] and
[O~II] has been taken into account. In order to simulate the lower limits on
temperature imposed by photo-ionization from the central star (without explicit
calculation of the photo-ionization) a lower limit of 5000~K has been kept
for the temperature. The grid size at full resolution is  $513\times257$
grid cells. The full domain
has a physical size of $1\times10^{18}$~cm by $5\times10^{17}$~cm. The outflow
was initialized on a sphere with a radius of $5\times10^{16}$~cm. We have used
a Courant-number of 0.1 times the smallest signal crossing-times combined with
the cooling time scale of all active cells on the adaptive grid.

The three-dimensional simulation has been produced with the Reefa-code
described in Lim \& Steffen (2001). This code uses a similar scheme as 
Coral (Raga et al. 1995) to solve the hydrodynamic equations on a binary 
adaptive 3D-grid. The grid size is variable and increases as the
expanding dynamical region nears the boundary (at the time shown, the simulation in
Figure 3d has dimensions of $345\times345\times289$ cells). Again we
use the flux-vector splitting method due to Van Leer (1982), with
the mass density conservation  equation written separately for each of the
ionic and neutral components.  The radiative energy loss is computed with the
prescription described by  Biro \etal (1995). The collisional ionization of
hydrogen by  electron impact is included with the rate coefficients of Cox
(1970) and  recombination with the interpolation formula of Seaton (1960).

\subsection{Initial and boundary conditions}

% 
% OBSERVED STRUCTURAL AND DYNAMICAL PROPERTIES TO BE REPRODUCED IN SIMULATIONS
%

The main observed dynamical and structural properties which served as
a guideline for our simulations are the following: radius of the
brightest sections of the nebula between 0.25 and 0.5 parsec, rim densities
of order 1,000\pccm\ with expansion speeds near 50\kms\ at this size.
The properties of the axial FLIERs should include a density of the order of
$10^4$\pccm\ and speeds between that of the rim and around
200\kms, which may rise linearly with distance up to velocities over
600\kms\ , as observed in the case of MyCn 18 (Bryce et al. 1997, O'Connor
et al. 2000). 

The initial density distribution $n(\theta)$ has been
assumed to be static varying with polar angle $\theta$ according to
the equation 
\begin{equation}
\label{environ.eq}
n(\theta,r) = n_0 \left[(1-q)\sin^\delta\theta +q\right] (r/r_0)^{-\alpha}.
\end{equation}

This is equivalent to the one used by Kahn \& West (1985),
except for the free power-law parameter $\alpha$. It takes into account 
possible deviations from the inverse square law, which might arise from
time-variations of the AGB-wind.  Here $n_0$ is the density at the fiducial
distance $r_0$ from the star, $q$ is the pole to equator density ratio,
$\delta$ controls how fast the density rises from pole to equator. This
density distribution in the environment fits calculations of stellar outflows
taking into account slow stellar rotation (Reimers, Dorfi, \& H\"ofner, 2000).
In the 3D-simulation the density of environment has been assumed to be
uniform.

The latitudinal momentum distribution in the fast wind will depend on the
model adopted for its origin. 
%
%SECOND REVISION 6
%
Precessing jets are one possibility to produce the
required momentum distribution for the formation of stagnation knots.
On the other hand, rather uncollimated interacting winds from double
stars are another possible mechanism to produce a deficiency of momentum
perpendicular to the plane of the binary (Soker \& Rappaport, 2000). However,
at this time it is not clear whether this sort of axial reduction of momentum
can hold up to the distances required for the formation of stagnation knots.
Currently single stellar wind scenarios do not seem to be able to 
produce polar momentum deficiencies. Considerable simplifications
in the theory of winds from rapidly rotating stars do, however,
leave some open ground for winds with polar momentum deficiencies 
(Petrenz \& Puls, 2000; S. Cranmer private communication).  Hence,
the details of the velocity distribution used in this paper are to be
regarded only as one possibility of many. Simulations with other distributions
show essentially the same qualitative results. 

Since there is no simple analytic description of the wind
momentum as a function of polar angle, for this initial qualitative study, we 
therefore adopt the velocity distribution $v(\theta)$ of the fast wind
(equation \ref{velocdist.eq}) to follow the same function of polar angle as
the density of the slow wind, with the difference that the maximum speed $v_0$
is reached at some half opening angle $\theta_0$.  At higher polar angles than
$\theta_0$ the wind has constant velocity.  The numerical parameters $\kappa$
and $\epsilon$ may of course have different values to $q$ and $\delta$ in
equation \ref{environ.eq}.  % 

\begin{equation} 
\label{velocdist.eq} 
\frac{v(\theta)}{v_0} = 
(1-\epsilon) \sin^\kappa \left(\frac{\pi \theta}{2\theta_0}\right)+\epsilon
\end{equation}

For the simulations we have used the parameters shown in Table 1.

\section{Results and Discussion}

We have performed a series of simulations varying the following parameters: the
pole-equator ratios of the velocity of the fast flow, the densities, 
the opening-angle of the zone of momentum depletion, and the pole-equator
density ratio in the ambient medium. Figure 1 shows representative
results.

In the simulations we identify the
three main structures which are observed in many PNe. First, a high density
envelope or rim of shocked external gas which propagates at a few tens of
kilometers per second (see also Figure 1). Second, small knots formed by
instabilities in the envelope, which might correspond to
low-ionization knots which are not aligned with the symmetry axis of the nebula
(Dwarkadas \& Balick 1998). Third, fast dense knots near the axis which we
associate with the ansae and which are the subject of this paper. 

As long as the central outflow is on, driving the expanding shock, the
stagnation knot moves roughly at the same speed as the rest of the
bow-shock and remains at the bright rim formed by shocked ambient gas.
As soon as the outflow ceases, however, the
cooling envelope slows down rather quickly, whereas the dense knot continues to
move at its original speed. Later, it slows down as it expands and continues
to add mass from the ambient medium. The deceleration of the
expanding envelope is ever so faster than that of the stagnation knot,
since the ambient density in the equatorial region is higher than
that in the polar region. 

Instabilities in the thin cold envelope (representing
the main PN-shell) may develop. In these simulations the development of
instabilities is susceptible to the details of the numerical procedure, e.g.
the Courant number, resolution and diffusion of the code. The kinematic
results on them are therefore not very reliable (see Dwarkadas and Balick,
1998, for a  discussion of the instabilities). The stagnation knot is not
noticeably sensitive to the numerical details, because it does not form on the
scale of the grid resolution.

Figures 3a and 3b show stagnation knots of the
slow (RUN 1) and the fast (RUN 2) types, respectively (see Table 1 for the
parameters). The slow knot starts to expand significantly before it separates
much from the rim of the nebula. Small deviations from the transverse
viewing angle will make the knot appear to be inside the PN-shell. Its speed at
the time of the density image shown in Figure 1a is around 140\kms. 
The fast knot in Figure 3b is able to reach a greater distance before it is
slowed down as a consequence of its expansion. The expansion causes an
increase of the  rate at which mass is swept up, and hence greater
deceleration. The speed of the fast stagnation knot reaches 240\kms\ at the
tip. 

% SECOND REVISION 1
Inspection of the structure of the axial FLIERs in the simulations reveal
different possible morphologies. We find head-tail
structures directed both ways, towards and away from the source, as those
observed in the HST images of NGC 3242, 6826 and 7009, presented by Balick et
al. (1998). Tails directed towards the central star are, however, short-lived
(see below). RUN 1 (Figure 3a) shows an extended blob, with some head-tail
structure, with the head towards the source. RUN 2 is similar to
the structures found in the Saturn-nebula (NGC~7009), where we find a crossing
shock feature on the axis with the highest density on the near side of the
star. This feature lies within the main rim. Outside the rim, there is the main
stagnation knot, which has developed a bow-shock structure, with highest
density near the tip. Lower density wings extend back towards the source.

The main differences in the conditions from which me obtain one or another
type of stagnation knot is as follows. The slow knot is produced in the case
where there is a strong reduction of momentum along the axis (by a factor of
0.2), a relatively high ambient density along the axis and a smaller wind
speed. The reverse is true for the fast knot. Here the reduction of momentum
is only moderate (by a factor of 0.5). Densities in both stagnation knots vary
roughly between $10^3$ and $10^4$\pccm.

Figure 3c (table 1 RUN 3) shows a very long stagnation knot or jet. Its nature
is very different from that of jets from young stellar objects or 
active galaxies, which are collimated very near the central source and are 
refueled continuously or in pulses. In the stagnation knots, 
we have swept-up material which is not replaced from the central source. The
most important parameters for the production of such a long jet are a small
opening angle of the  region of reduced momentum flow and a large pole-equator
density ratio in  the environment. 

%SECOND REVISION 2
The axial jet-like structure initially has its highest density on the
far-side as seen from the star, since this is where catastrophic cooling
starts first. At that time, this feature would be seen as a head-tail
structure with the head and lowest-ionization on the far side. This situation
is, however, transient (about 100 years in this simulation). After this time,
the reverse is true, i.e. the denser region with presumably lower ionization
faces the star. The reason for this is that the far side collapses first and
reexpands first as well, such that later it has lower density as compared
to the region closer to the centre. 

We find that in the stagnation "jets", the velocity 
increases linearly with distance  (see inset of Figure 3c). This velocity
structure is very similar to that seen in the  magneto-hydrodynamic models of
ansae developed by Garc\'{\i}a-Segura et al. (1999). Both models, the
stagnation knots and the MHD-model reproduce this important observational
result. This problem deserves a separate study. 

The three-dimensional calculations (Figure 3d, table 1 RUN 3D) show that the
formation of stagnation knots is not a consequence of the symmetry imposed
in the axisymmetric calculations. In 3D, the knots form and develop in a very
similar fashion to 2D simulations.

%SECOND REVISION 3
From our simulations we can discuss the ionization state of the knots only
as  suggested by the density structure, assuming ionization balance with
respect to incoming photons. Thus to first order we take that high density in
the knots implies  low ionization and vice-versa. A detailed ionization study
will have to include not only photo-ionization from the central star, but also
collisional and possibly local photo-ionization from strong and fast shocks, 
since these are ubiquituous in our model. These local sources of ionization
may complicate the ionization structure considerably.
Such a study is outside the scope of the present paper.

%SECOND REVISION 4
As for the geometry of the polar region predicted by the stagnation
knot model, a comparison with observations has to focus on young PNe, 
since the expected concavity (in the extended definition proposed here) is only
short-lived (of the order of a few hundred years).  Hence, we expect to find
the concavity  only in a rather small fraction of objects in which, in
addition, the axial FLIERs have not yet fully developed. The
young PN IRAS~17150-3224 is such an example. It shows a pair of elongated,
truncated lobes with concave ends (Kwok, Su \& Hrivnak, 1998). An additional
clear example of a PN showing a concave bow-shock structure is K 3-24
(Manchado et al. 1996). This type of structure is best obtained with a conical
momentum distribution, i.e. with a clear maximum at some polar angle (Steffen
\& L\'opez, 2000). Physical mechanism for producing such a momentum
distribution can be, e.g., a rapidly precessing jet (Peter \& Eichler, 1995;
Lim 2000) or a close binary system where one of the components blows a
collimated fast wind (Soker \& Rappaport, 2000).

\section{SUMMARY}

We have shown that the high-velocity axial FLIERs found in 
some bipolar/elliptical planetary nebulae could be due to a polar deficiency
of the momentum flow of the fast stellar wind which interacts with the
ambient medium. This wind configuration accumulates and compresses
shocked {\em ambient} gas in the stagnation region of the concave (locally
inverted) bow-shock. We find basically three different types of stagnation
features. First, those which remain on or within the rim of the PN. Second,
knots which escape the rim and, third, long jet-like features which may
disintegrate to form a series of knots. The stagnation knots can be produced
with speeds of up to several hundred kilometers per second. The stagnation
jets show a constant positive velocity gradient with distance from the source.
Both the magnitude and the gradients of the velocities are consistent with
observations.

\begin{acknowledgements}
W.S. is grateful for travel support from project DGAPA-UNAM IN114199
and University College London visiting researchers programme. 
J.A.L. acknowledges support from project DGAPA-UNAM IN114199. 
A.J.L. acknowledges a PPARC research associateship. We thank Guillermo 
Garc\'{\i}a-Segura and Alejandro Raga for useful discussions. 
We are also greatful to the constructive suggestions by the anonymous
referee.
\end{acknowledgements}

\begin{figure}[ht]
\plottwo{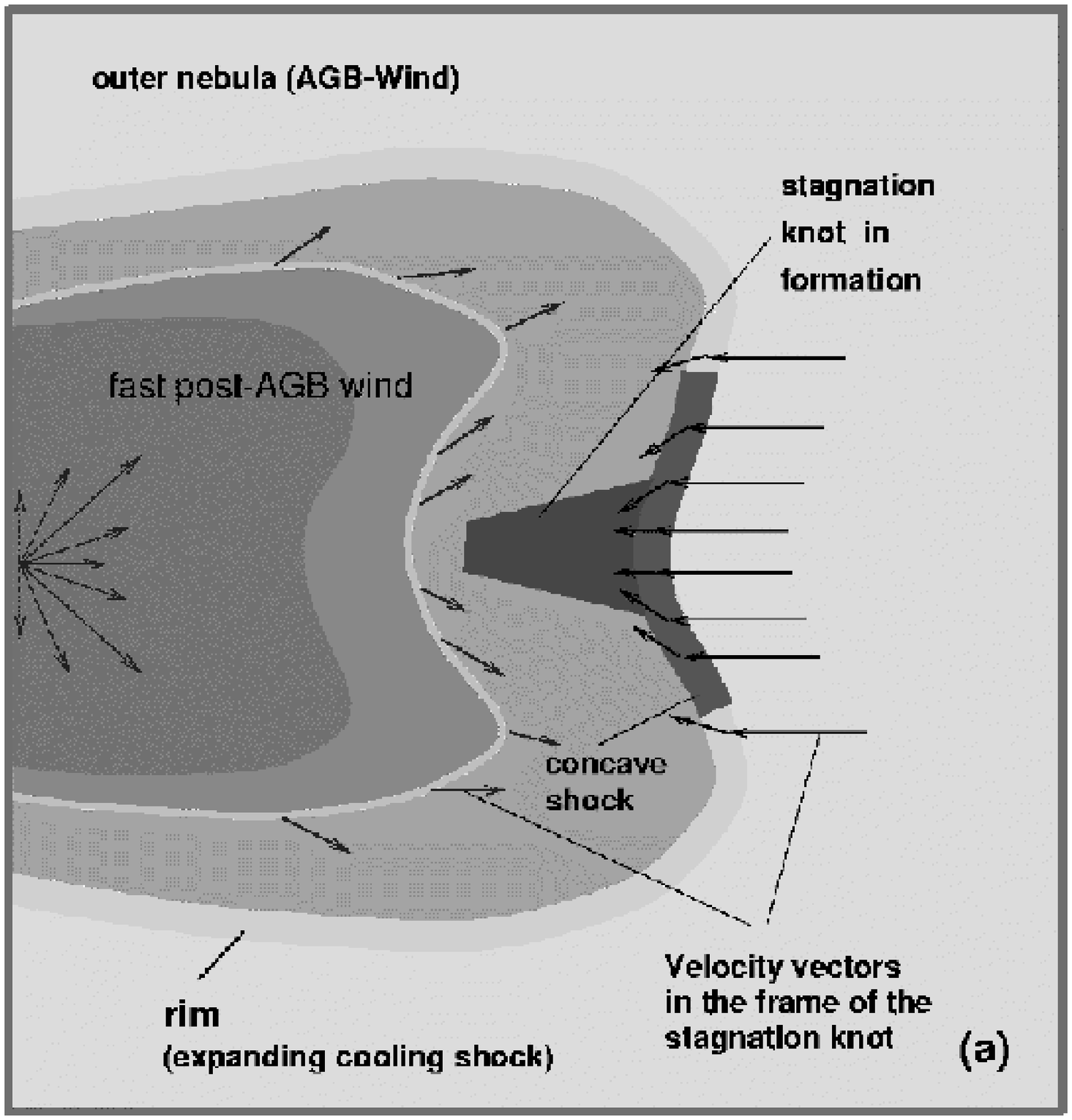}{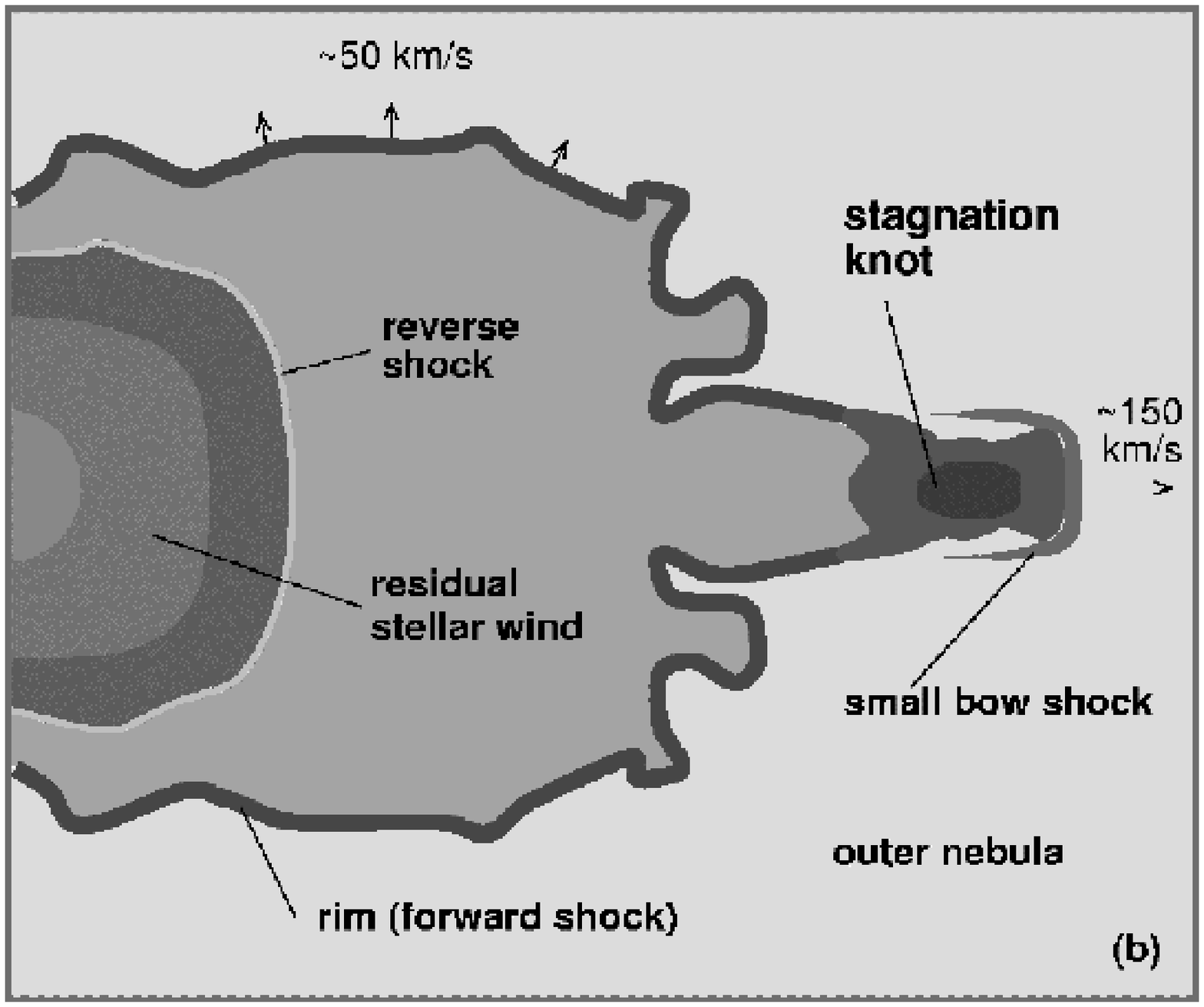}
%\plotone{stagn_sch3_grey.ps}
%\plotone{stagn_sch2_grey.ps}
\caption{
Schematic view of the early evolution of stagnation knot during the
phase of a concave bow-shock (Panel a). The arrows near the origin represent
the  moment flux in the fast wind and are drawn in the rest frame of the
central star. The arrows in the bow-shock region are velocities of 
the shocked fast wind (left) and the shocked external medium (right) 
{\em in the rest frame of the stagnation knot}. They illustrate
the divergent flow of the fast wind and the {\em convergence} of the 
slow wind. Panel b illustrates the late stage of evolution during which the
stagnation feature has separated from rim of the planetary nebula. }  
\end{figure}

\begin{figure}[ht]
\plotone{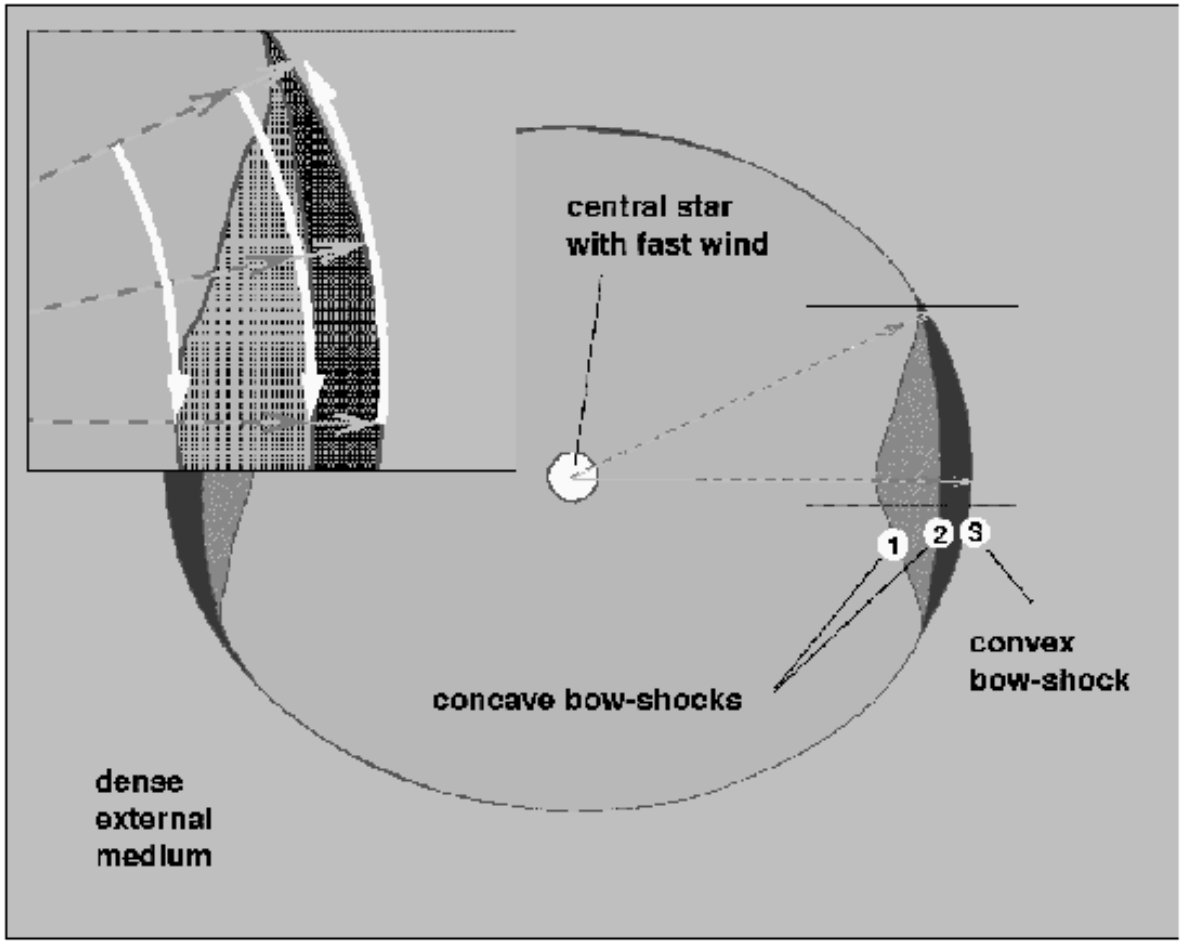}
\caption{
Schematic view of an elliptical shock wave of the fast stellar wind 
moving into the external medium with different bow-shock structures 
near the symmetry axis. Three qualitatively different cases are shown. 
Case 1 is a clearly concave bow-shock. Case 2 is geometrically convex or
almost planar, but with respect to the spherical wind it acts as a concave
shock (see inset were the position of the shock wave is shown with respect to
the flow lines of the stellar wind).  Case 3 is the normal convex bow-shock.
The white arrows in the inset indicate the tangential lines to the stellar wind
and the directions  of the pressure gradient produced by the shocked fast
stellar wind. }   \end{figure}

\begin{figure}
\plotone{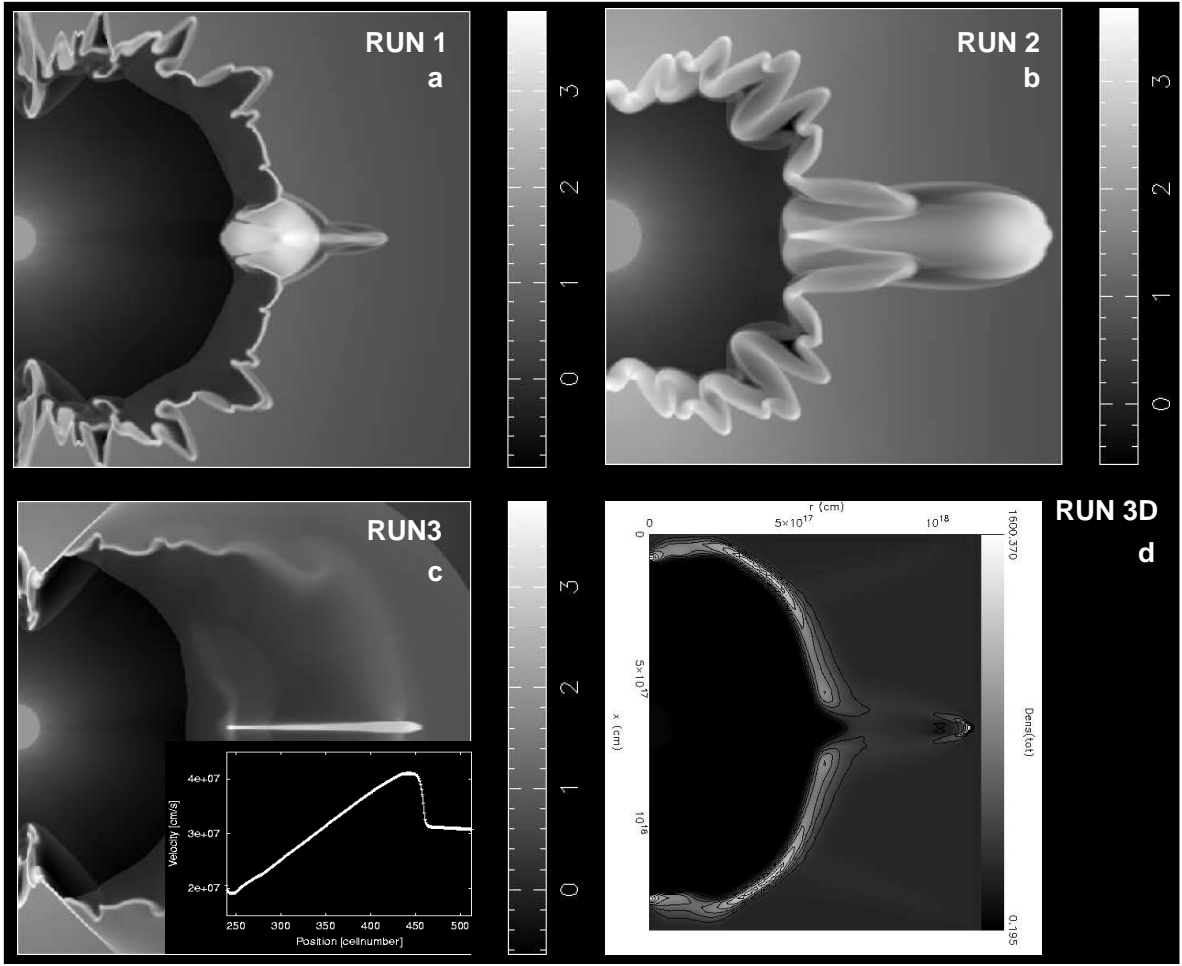}
\caption{
Logarithmically scaled density cuts through axisymmetrical simulations are
shown. Panel "a" shows a stagnation knot which expands significantly before
escaping the rim of the PN, whereas in RUN 2 of panel "b" it propagates about
twice as far before reaching a similar expansion and has escaped the main PN. 
In panel "c" a "butterfly"-type PN is formed with a stagnation jet along the
symmetry axis. The inset is a plot of the velocity of the gas on
the  axis given in units of cm/s. Note the linear increase in velocity as a
function of distance in the region of the dense axial jet. Panel "d" shows
the 3D-simulation with the stagnation knot escaping the rim of the PN in a
similar fashion as it does in the cylindrical 2D-runs. 
}  
\end{figure}

\begin{table}
          \begin{tabular}{lllll}
            \hline            & RUN 1 & RUN 2 & RUN 3 & RUN 3D \\
            \hline
            \hline Time       & 1190  &  1164 & 604  & 4053  \\
            \hline WIND       &       &       &      &       \\
            \hline $v_0$      & 1000  &   700 & 1000 & 1000  \\
            \hline $n_0$      & 50    &    50 &   50 &   40  \\
            \hline $\epsilon$ & 0.5   &   0.2 &  0.5 &  0.5  \\
            \hline $\kappa$   & 2     &     2 &    2 &    2  \\
            \hline $\theta_0$ & 50    &    30 &   20 &   20  \\
            \hline
            \hline ENV.     &      &       &      &      \\
            \hline $n_0$    & 4000 &  2000 & 3500 & 240  \\
            \hline $r_0$    & 8    &     5 &    5 &   8  \\
            \hline $q$      & 0.2  &   0.5 &  0.2 &   1  \\
            \hline $\delta$ & 4    &     4 &    4 &   4  \\
            \hline $\alpha$ & 1.95 &  1.95 & 1.95 &   0  \\
            \hline
           \end{tabular}  
\caption{
The model parameters are given as used in equations \ref{environ.eq} and
\ref{velocdist.eq}. The units are as follows: time in years, velocity $v$ in
\kms, density $n_0$ in amu\pccm, the angle $\theta$ in degrees, and the
fiducial radius $r_0$ in units of $10^{16}$cm. Other quantities have no units.
The cut-off time of the stellar wind is $4\times10^9$\,sec in all cases.  
}
\end{table}


\begin{thebibliography}{}
%
\bibitem{}
Aller, L.H. 1941, ApJ, 93, 236 
% Name "ansae" for symmetrical microstructures of PNe
%
\bibitem{}
Balick, B., Rugers, M., Terzian, Y., Chengalur, J. N. 1993, 
ApJ, 411, 778 
% ground-based observations of FLIERs 
% 
\bibitem{}
Balick, B., Alexander, J., Hajian, A.R., Terzian, Y., Perinotto, M.,
Patriarchi, P. 1998, AJ, 116, 360   
% High-res. observations of FLIERs 
%
\bibitem{}
Biro, S., Raga, A.C., Cant\'o, J. 1995,
MNRAS, 275, 557 
% Cooling in code
%
\bibitem{}
Bobrowsky, M., Sahu, K. C., Parthasarathy, M., Garc\'{\i}a-Lario, P. 1998,
Nature, 392, 469 
% Birth and early evolution of a planetary nebula.
%
\bibitem{}
Bryce, M., L\'opez, J.A., Holloway, A.J., Meaburn, J. 1997,
ApJ, 487, 161 
% MyCn18 velocity gradient
%
\bibitem{}
Dopita, M.A. 1997, 
ApJ, 485L, 41 
%
\bibitem{}
Cant\'o, J., Tenorio-Tagle, G., R\'o$\dot{\rm z}$yczka, M. 1988,
A\&A, 192, 287 
% Cant\'o mechanismo for jet collimation
%
\bibitem{}
Cox, D.P., 1970,
PhD Thesis (University of California, San Diego) 
% rate coefficients
%
\bibitem{}
Dwarkadas, V.V.; Balick, B. 1998, ApJ, 497, 267 
% Instabilites in PN-simulations as FLIERs
%
\bibitem{}
Frank, A., Balick, B., Livio, M. 1996, ApJ, 471L, 53 
% Canto mechanism for FLIERs
%
\bibitem{}
Garc\'{\i}a-Segura, G., Langer, N., R\'o$\dot{\rm z}$yczka, M. 1999, 
ApJ, 517, 767 
% Shaping Bipolar and Elliptical Planetary Nebulae: 
% Effects of Stellar Rotation, Photoionization Heating, and Magnetic Fields
%
\bibitem{}
Garc\'{\i}a-Segura, G., L\'opez, J.A. 2000,
ApJ, 544, 336  
% Three-dimensional Magnetohydrodynamic Modeling of Planetary Nebulae. II.
%
\bibitem{}
Goncalves, D.R., Corradi, R.L.M., Mampaso, A. 2000,
ApJ, in press
% Low-ionization structures in planetary nebulae:
% confronting models with observations
%
\bibitem{}
Guerrero, M. A., V\'azquez, R., L\'opez, J. A. 1999,
AJ, 117, 967
% The Kinematics of Point-symmetric Planetary Nebulae
%
\bibitem{}
Jones, T.W., Kang, H., \& Tregillis, I.L. 1994, ApJ, 432, 194  
% Hydro of bullets in low-density medium
% 
\bibitem{}
Kahn, F., West, K. 1985,
MNRAS, 212, 837
%
\bibitem{}
Kwok, S., Su, K.Y.L., Hrivnak, B.J. 1998,
ApJ, 501, L117
% HST V-Band imaging of PPN IRAS 17150-3224
% 
\bibitem{}
Lim, A. 2001,
MNRAS, submitted. 
% stagnation knots in precessing jets
%
\bibitem{}
Lim, A., Steffen, W. 2001,
MNRAS, 322, 166
% Reefa, 3D adaptive code
%
\bibitem{}
L\'opez, J.A. 2000,
RMxAC, 9, 201 
% Collimated Outflows in Planetary Nebulae
%
\bibitem{}
Manchado, A., Guerrero, M., Stanghellini, L., Serra-Ricart, M. 1996,
The IAC Morphological Catalog of Northern Galactic Planetary Nebulae,
(La Laguna: IAC) 
%
\bibitem{}
Meaburn J.,  Lopez, J.A., Noriega-Crespo, A. 2000, 
ApJSS, 128, 321 
% Halos
%
\bibitem{}
Moreno-Corral M.A, de la Fuente, E., Gutierrez, F. 1998,  
RevMexAA 34, 117. 
% NGC7009
%
\bibitem{}
O'Connor, J.A., Redman, M.P., Holloway, A.J., Bryce, M.,
L\'opez, J.A., Meaburn, J., 2000, ApJ, 531, 336  
% Observations of MyCn18, line of fliers with linear increase 
% of speed up to 500 km/s 
%
\bibitem{}
Pe\~na, M., Hamann, W.-R., Koesterke, L., Maza, J.,
 Mendez, R. H., Peimbert, M., Ruiz, M. T., Torres-Peimbert, S. 1997,
ApJ, 491, 233 
% Spectrophotometric Data of the Central Star of the 
% Large Magellanic Cloud Planetary Nebula N66: ...
%
\bibitem{}
Peter, W., Eichler, D., 1995,
ApJ, 438, 244 
% Inertial confinement of astrophysical jets
%
\bibitem{}
Petrenz, P., Puls, J. 2000,
A\&A, 358, 956
% ... radiation driven winds from rotating early-type stars
%
\bibitem{}
Raga, A.C., Taylor, S. D., Cabrit, S., Biro, S. 1995, 
MNRAS, 296, 833  
% Code
%
\bibitem{}
Reay, N.K., Atherton, P.D. 1985, 
MNRAS, 215, 233    
% Observations of NGC7009
%
\bibitem{}
Redman, M.P., Dyson, J.E. 1999, MNRAS, 302, L17  
% FLIERs as recombination fronts in mass-loading jets
%
\bibitem{}
Reimers, C., Dorfi, E.A., H\"ofner, S. 2000,
A\&A, 354, 573 
% Shaping of elliptical planetary nebulae. 
% The influence of dust-driven winds of AGB stars
%
\bibitem{}
R\'o$\dot{\rm z}$yczka, M., \& Franco, J. 1996, ApJ, 469, L127
% Initial MHD numerical model of collimated stellar wind
%
\bibitem{}
Seaton, M. J. 1960,
Report on Progress in Physics, 23, 313 
%
\bibitem{}
Soker, N., Rappaport, S. 2000, 
ApJ, 538, 241  
% Interacting winds in PN
%
\bibitem{}
Soker, N., Regev, O. 1998, AJ, 116, 2462  
% Bullets as FLIERs
%
\bibitem{}
Steffen, W., L\'opez, A.J. 1998, ApJ, 508, 696  
% Model KjPn8
%
\bibitem{}
Steffen, W., L\'opez, A.J. 2000,
Asymmetrical Planetary Nebulae II: From Origins to Microstructures, 
ASP Conf. Series, Vol. 199, p. 413
Eds. J. H. Kastner, N. Soker, and S. Rappaport 
% ansae as stagnation knots
%
\bibitem{}
van Leer, B. 1982, 
in Lecture Notes in Physics 170, Numerical Methods in Fluid Dynamics, 
ed. E. Krause, (Berlin: Springer), 507
%
\bibitem{}
Werner, K., Hamann, W.-R., Heber, U., Napiwotzki, R.,
Rauch, T., Wessolowski, U. 1992,
A\&A, 259L, 69  

\end{thebibliography}
\end{document}